\documentclass{PoS}
\newcommand{\mbf}{\mathbf}

\newcommand\lsim{\mathrel{\rlap{\lower4pt\hbox{\hskip1pt$\sim$}}
    \raise1pt\hbox{$<$}}}
\newcommand\gsim{\mathrel{\rlap{\lower4pt\hbox{\hskip1pt$\sim$}}
    \raise1pt\hbox{$>$}}}

\newcommand{\mrm}{\mathrm}

\title{Status of chiral perturbation theory}

\ShortTitle{Status of chiral perturbation theory}

\author{\speaker{Gerhard Ecker}\thanks{Supported in part by EU contract
No. MRTN-CT-2006-035482 (FLAVIAnet)}\\
Faculty of Physics, University of Vienna, Boltzmanngasse 5, A-1090 Wien \\
        E-mail: \email{gerhard.ecker@univie.ac.at}}

\abstract{The status of chiral perturbation theory in the
  meson sector is reviewed. The main emphasis is on recent
  developments in pion pion scattering, in semileptonic decays and in 
  nonleptonic kaon decays. A few other selected topics are also
  discussed.}

\FullConference{8th Conference Quark Confinement and the Hadron Spectrum \\
		 September 1-6, 2008\\
		 Mainz, Germany}

\begin{document}

\section{Introduction and overview}
\noindent 
Chiral perturbation theory (CHPT) was introduced nearly 30 years ago
\cite{weinberg79}. Soon thereafter, it was developed
\cite{glsu2,glsu3} into a powerful tool for treating the strong 
interaction at low energies ($E \ll 1$ GeV). From the first steps
beyond current algebra, CHPT has come a long way as manifested by the
present-day chiral Lagrangian in the meson sector\footnote{The
baryon sector is covered by Mei\ss ner \cite{ulfmainz}.} displayed in
Table \ref{tab:effmL}. In addition to the 
strong interactions of mesons, the Lagrangian also accounts for
nonleptonic weak interactions and it includes the photon and 
leptons as dynamical degrees of freedom. CHPT aims at reliable results 
in the confinement regime to allow for conclusive tests of the
Standard Model and to search at the same time for traces of new
physics at low energies.

\begin{table}[h]
$$
\begin{tabular}{|l|c|} 
\hline
&  \\[-.2cm] 
\hspace{1cm} ${\cal L}_{\rm chiral\; order}$ 
~($\#$ of LECs)  &  loop  ~order \\[8pt] 
\hline 
&  \\
 ${\cal L}_{p^2}(2)$~+~ 
${\cal L}_{p^4}^{\rm odd}(0)$
~+~ ${\cal L}_{G_Fp^2}^{\Delta S=1}(2)$  
~+~ ${\cal L}_{G_8e^2p^0}^{\rm emweak}(1)$ & $L=0$
\\[3pt] 
~+~ ${\cal L}_{e^2p^0}^{\rm em}(1)$ ~+~
${\cal L}_{\mrm{kin}}^{\rm leptons}(0)$ & \\[10pt]
~+~ ${\cal L}_{p^4}(10)$~+~
${\cal L}_{p^6}^{\rm odd}(23)$
~+~${\cal L}_{G_8p^4}^{\Delta S=1}(22)$
~+~${\cal L}_{G_{27}p^4}^{\Delta S=1}(28)$ &   
$L \le 1$ \\[3pt]
~+~${\cal L}_{G_8e^2p^2}^{\rm emweak}(14)$ 
~+~ ${\cal L}_{e^2p^2}^{\rm em}(13)$
~+~ ${\cal L}_{e^2p^2}^{\rm leptons}(5)$  & \\[10pt] 
~+~ ${\cal L}_{p^6}(90)$  & $L \le 2$ \\[8pt] 
\hline
\end{tabular}
$$
\caption{The mesonic Lagrangian for chiral $SU(3)$ in use today,
  including strong, electromagnetic and nonleptonic weak
  interactions. The leptons must be incorporated for radiative 
  corrections in semileptonic decays. The numbers
  in brackets denote the number of low-energy constants (LECs).}
\label{tab:effmL}
\end{table}

As Table \ref{tab:effmL} shows, higher orders in the chiral expansion
are accompanied by an increasing number of
LECs. The determination of those LECs has been and will continue to be
essential for progress in the field. The following methods are being
employed.
\begin{itemize} 
\item[i.] The LECs can be determined by confronting CHPT predictions 
with experimental data. This straightforward approach
  encounters its limits already at NLO for
  nonleptonic weak decays and at NNLO for 
  strong processes: there are too many LECs for the experimental
  information available to obtain a predictive scheme. 
\item[ii.] In some cases, combining chiral amplitudes
  with dispersion theory has proven to be fruitful.
\item[iii.] Lattice methods have made tremendous progress 
\cite{necco,heirimainz}.
\item[iv.] Large-$N_c$ motivated resonance saturation has provided a
  number of successful estimates \cite{toni}.
\end{itemize} 

The purpose of this talk was to discuss some major achievements
of CHPT in the meson sector during the past two years: pion pion
scattering, semileptonic decays, nonleptonic $K$ decays and a
few other selected topics.

\section{Pion pion scattering}
\noindent 
Pion pion scattering at low energies is a fundamental process for
chiral $SU(2)$. It is in particular very sensitive to the mechanism of 
spontaneous chiral symmetry breaking. The ultimate theoretical result
was obtained by combining dispersion theory \cite{acgl01} with
CHPT. The  dispersive amplitude 
depends on experimental data for $E_{\pi\pi} \ge 800$ MeV and on two
subtraction constants. By matching the dispersive amplitude with the 
chiral amplitude to NNLO \cite{pipiop6}, the 
two subtraction constants can be expressed in terms of the S-wave
scattering lengths \cite{cgl01}
\begin{eqnarray}
a^0_0=0.220\pm 0.005~, \hspace*{1cm} & \hspace*{1cm} 
a^2_0 = - 0.0444 \pm 0.0010 ~.
\label{eq:a0a2}
\end{eqnarray}
When comparing theoretical predictions with experimental results,
one must keep in mind that the chiral amplitudes and the scattering
lengths (\ref{eq:a0a2}) refer to an isospin symmetric world. Isospin
violation and radiative corrections must therefore be taken into
account before making the comparison.

\noindent 
Experimental information comes from three sources.
\begin{itemize} 
\item[a.] $K_{e4}$ decays;
\item[b.] Decay of pionium; 
\item[c.] Cusp in $K \to 3 \pi$ ~decays.
\end{itemize} 
I will restrict the discussion to items a) and c) where important
developments, both in experiment and in theory, have occurred during
the last two years.

\subsection{$\mbf{K^+ \to \pi^+ \pi^- e^+ \nu_e}$ decays}
\noindent 
$K_{e4}$ decays are the traditional source for accessing 
pion pion scattering at low energies. From the final state interaction
of the two pions one can extract the phase shift difference $\delta_0
- \delta_1$ where  $\delta_0, \delta_1$ are the phase shifts for the
$I=0$ S-wave and the $I=1$ P-wave, respectively, in the isospin limit
and in the absence of electromagnetic corrections.

\begin{figure}[ht]
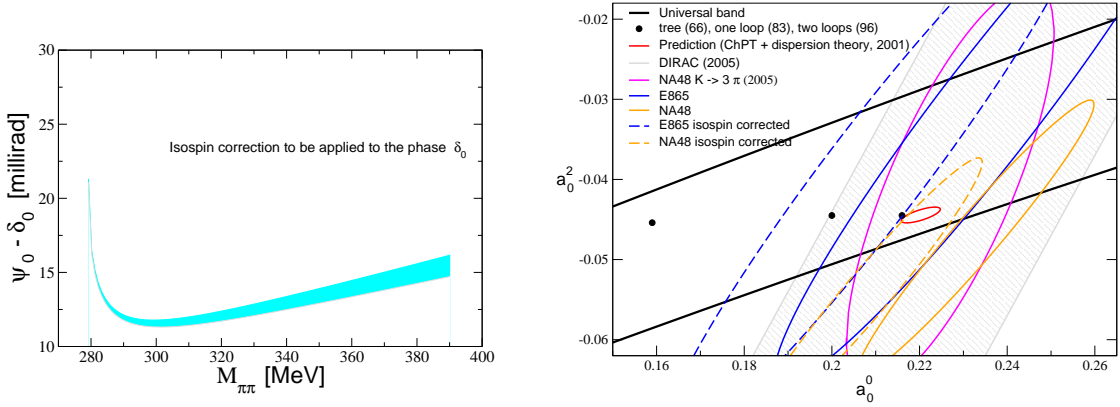

\setlength{\unitlength}{1cm}
\begin{picture}(16,8)
\put(0.1,0.03){\makebox(6.5,7.0)[lb]{
\leavevmode 
\includegraphics[width=6.5cm]{phasediff_ke4}}}
\put(7.5,-0.1){\makebox(7.5,7.5)[lb]
{\leavevmode 
\includegraphics[width=7.5cm]{a0a2_paper}}}
\end{picture}
\caption{Isospin violating corrections to be applied to the measured
  S-wave phase shift (left figure). Comparison between theory
  and experiment for the scattering lengths (right figure). Both
  figures are taken from Ref.~\cite{cgr2008}.} 
\label{fig:a0a2}
\end{figure}

In addition to the radiative corrections applied by the experimental
groups, isospin violation due to the pion mass difference
and to $m_u - m_d$ has turned out to be important. The most recent
analysis at the one-loop level is described in Ref.~\cite{cgr2008}
where also references to related previous and ongoing work can be
found.  In the one-loop diagrams
for $K_{e4}$, the physical $\pi^+,\pi^0$ masses must be inserted and
an additional diagram involving $\pi^0 - \eta$ mixing appears. Denoting
the experimentally accessible phase shifts by $\psi_0,\psi_1$, the
authors of Ref.~\cite{cgr2008} obtain for the measurable S-wave phase
shift $\psi_0(s)$ in the elastic region $4 M_{\pi^+}^2 < s < 16
M_{\pi^0}^2$  ($\psi_1=\delta_1$ to the order considered)
\begin{equation}
\psi_0(s) = \displaystyle\frac{1}{32\pi F^2} \left\{(s + 4 M_{\pi^+}^2 -
  4 M_{\pi^0}^2)\sigma(M_{\pi^+}) + (s - M_{\pi^0}^2) \left(1 +
  \displaystyle\frac{3(m_d - m_u)}{2(m_s - \hat{m})}  
  \right)\sigma(M_{\pi^0}) \right\} + O(p^4)
\end{equation}
where
\begin{eqnarray}
\sigma(M) = \sqrt{1 - 4 M^2/s}~, \qquad  & \qquad \hat{m} = \frac{1}{2}
(m_u + m_d) ~. 
\end{eqnarray} 
The difference $\psi_0 - \delta_0$ to be subtracted from the measured
phase shift $\psi_0(s)$ is shown in Fig.~\ref{fig:a0a2}
\cite{cgr2008}. The isospin
  corrected scattering lengths from the NA48 experiment at CERN
  \cite{NA48ke4} now agree perfectly with the theoretical prediction
  \cite{cgl01} (the small red ellipse in Fig.~\ref{fig:a0a2})
  whereas the agreement is now less impressive for the BNL experiment
  E865 \cite{E865ke4}. For further details I refer to the contribution
  of Bloch-Devaux in these Proceedings \cite{bbd}. 

\vskip -2.5 cm

\subsection{Cusp in $\mbf{K \to 3 \pi}$ decays}
\noindent 
The cusp was first seen in the $M_{\pi^0\pi^0}$ distribution in 
$K^\pm \to \pi^\pm \pi^0 \pi^0$ \cite{na48cusp}, more recently also in
$K_L \to 3 \pi^0$. It is due to the rescattering of pions in the final
state \cite{cabibbo04,mms97}
\begin{equation}
K^\pm \to \pi^\pm (\pi^+ \pi^-)^* \to \pi^\pm \pi^0 \pi^0 ~.
\label{eq:cusp}
\end{equation}
The basic mechanism is an interference between tree and 1-loop
amplitudes. The square-root singularity generates a cusp above
threshold at $M_{\pi^0\pi^0}^2 = 4 M^2_{\pi^+}$. From (\ref{eq:cusp})
the effect is seen to be sensitive to the combination of $\pi\pi$
scattering lengths
\begin{equation} 
a^0_0 - a^2_0 \sim 
A(\pi^+ \pi^- \to \pi^0 \pi^0)_{\mathrm{thresh}} ~.
\end{equation}  
Various approaches have been pursued to extract $a^0_0 - a^2_0$ from
$K \to 3 \pi$ near threshold.
\begin{enumerate} 
\item[i.] Following the original approach of Cabibbo \cite{cabibbo04}
  based on unitarity and analyticity, 
a systematic expansion of the singular terms of the $M_{\pi^0\pi^0}$
distribution in powers of the scattering lengths was performed in
Ref.~\cite{ci05}.
\item[ii.] In a related method, unitarity and analyticity were combined
  with CHPT \cite{gps07}.
\item[iii.] A two-loop dispersive representation of $K \to 3 \pi$
  amplitudes in the presence of isospin breaking is under construction
  \cite{kknz08}.  
\item[iv.] In the most advanced approach based on a nonrelativistic
  effective field theory, the $K \to 3 \pi$ amplitudes are expanded in
  powers of the scattering lengths and of the pion  momenta in the $K$
  rest frame \cite{cgkr06}. Most recently, radiative corrections have
  been performed within this framework \cite{bissetal08}. In contrast
  to standard CHPT, this approach is valid to all orders in the quark
  masses. 
\end{enumerate} 
An up-to-date comparison between theory and experiment can be found in
the contribution of Bloch-Devaux \cite{bbd}.

\section{Semileptonic decays}
\noindent 
Semileptonic decays have long been a rich field for CHPT
\cite{bcegdafne}.
\subsection{$\mbf{K_{l3}}$ decays}
\noindent 
$K_{l3}$ decays are at present still the best source for the CKM matrix
element $V_{us}$. They have therefore been investigated intensively
during recent years \cite{fkwg}. A possible problem with the slope of
the scalar form factor $f_0(t)$ was discussed by Leutwyler
\cite{heirimainz}. For the experimental analysis, the complete
radiative corrections in a CHPT framework  are now available for both
$K_{e3}$ and $K_{\mu 3}$ \cite{cgn08}. 
The collaboration between theory and experiment has led to a 
very precise value \cite{fkwg} for the product
\begin{equation} 
|V_{us}| f_+^{K^0\pi^+}(0) = 0.21661(47)~.
\end{equation}

\begin{figure}[ht]
    \vskip -.2 cm
    \begin{center}
    \begin{minipage}[t]{0.5\linewidth}
\vskip -5 cm

The predictions for the vector form factor 
at $t=0$ show a certain
spread, but are dominated now by the lattice results (see
Fig.~\ref{fig:fp0}). With $f_+^{K^0\pi^+}(0)=0.964(5)$ \cite{rbcuk08}
one obtains
$|V_{us}| =  0.2246(12)$~,
in perfect agreement with CKM unitarity, taking $|V_{ud}|$ from nuclear
$\beta$ decay. 
     \caption{Various predictions for $f_+^{K^0 \pi^+}(0)$ collected in 
Ref.~\cite{fkwg}.
     \label{fig:fp0} }
    \end{minipage}
    \hspace*{\fill}
    \begin{minipage}[t]{0.45\linewidth}
\begin{flushright} 
    \includegraphics[width=0.95\linewidth]{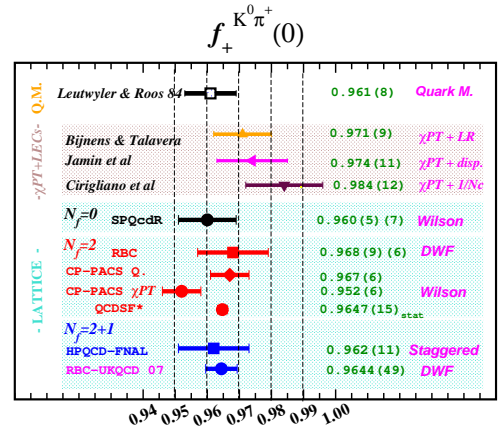}
\end{flushright} 
    \end{minipage}
   \end{center}
\end{figure}

\vskip -.2 cm
\noindent
For the sake of playing the devil's advocate, two
caveats are worth mentioning.
\begin{itemize} 
\item Hadronic $\tau$ decays are becoming competitive for the
  determination of $|V_{us}|$. Taking into account (small) $SU(3)$ 
breaking effects, Gamiz et al. obtained \cite{gamizetal07}
\begin{equation}
|V_{us}| = 0.2165(26)_{\rm expt}(5)_{\rm theor}~,
\label{eq:tau}
\end{equation}
in agreement with a similar more recent analysis of Maltman et
al. \cite{maltmanetal08}. The full data samples from BELLE and BaBar are
needed for a definitive conclusion. 
\item The most recent value $|V_{ud}|  = 0.97408(26)$ from
  superallowed $0^+ \to 0^+$ nuclear $\beta$ decay \cite{eronen08} has 
an uncertainty that is
  only half the theoretical uncertainty in pionic $\beta$ decay
  \cite{cknp03}, a much simpler process from the theorist's point of
  view. In this connection, I recall that a recent measurement
  of the 
neutron lifetime \cite{serebrov05} suggests a substantially bigger
$|V_{ud}|$, which together with CKM unitarity would imply a value for
$|V_{us}|$ even smaller than the $\tau$ decay result
(\ref{eq:tau}). However, the neutron lifetime from
Ref.~\cite{serebrov05} is incompatible with the present world average
\cite{pdg08}. 
\end{itemize}

\subsection{$\mbf{P_{l2}}$ decays  ~($\mbf{P=\pi, K}$)}
\noindent 
Because of the $V - A$ structure of charged currents the ratios 
$R_{e/\mu}^{(P)} =  \Gamma( P \to e \nu_e [\gamma] )/ 
\Gamma( P \to \mu \nu_\mu [\gamma] )$ are helicity suppressed. In
order to serve as sensitive probes for new physics, the Standard Model
values for these ratios must be known as precisely as possible. The
current experimental values are \cite{pdg08,fkwg}
\begin{eqnarray}
R_{e/\mu}^{(\pi)} = 1.230(4) \cdot 10^{-4}~, \qquad & \qquad 
R_{e/\mu}^{(K)} = 2.457(32) \cdot 10^{-5} ~. 
\label{eq:remuexp}
\end{eqnarray}
Turning off electromagnetic corrections, the ratios are given by the
classic values
\begin{equation}
R_{e/\mu}^P  =
\displaystyle\frac{m_e^2}{m_\mu^2}  \left(  \frac{M_P^2 - m_e^2}{M_P^2
  - m_\mu^2}   \right)^2
\end{equation}
to all orders in the chiral expansion. Nontrivial structure dependent
effects appear only for $e \neq 0$.

The corrections of $O(e^2 p^2)$ correspond to a point-like
approximation. The first systematic calculation to $O(e^2 p^4)$,
sensitive to the meson structure (form factors), was recently
performed by Cirigliano and Rosell \cite{ciro07}. 
The setup is the Lagrangian of Table \ref{tab:effmL} including
photons and leptons. In fact, the 2-loop calculation requires the
determination of a LEC of the Lagrangian of $O(e^2 p^4)$ that is not
included in Table \ref{tab:effmL} because the complete Lagrangian is 
not yet available. The relevant LEC is obtained by matching the
relevant form  factors with large-$N_c$ QCD \cite{ciro07}. The
associated uncertainty is accounted for in the final error
estimate. In contrast, the chiral double logs are model independent.

\begin{table}[ht]
\centering
\renewcommand{\arraystretch}{.8}
\begin{tabular}{|c||c|c|c|}  \hline 
 & & &    \\[-6pt]
 &  Cirigliano, Rosell \cite{ciro07}  & Marciano, Sirlin \cite{marsir93} 
& Finkemeier \cite{finke96}  \\[4pt] 
\hline
& &  &   \\[-8pt]
  $R_{e/\mu}^{(\pi)} \cdot 10^4$  &  $1.2352 \pm
 0.0001$ & $1.2352 \pm 0.0005$
&  $1.2354 \pm 0.0002$   \\[.3cm] 
 $R_{e/\mu}^{(K)} \cdot 10^5 $  & $2.477 \pm 0.001$ &   &
$2.472 \pm 0.001$  \\[-.4cm] 
  & & & \\
\hline
\end{tabular}
\caption{Theoretical predictions for $R_{e/\mu}^{(P)}$.}
\label{tab:remu}
\end{table}

Including photon emission and summing up the leading logs 
$\alpha^n \log^n (m_\mu/m_e)$ \cite{marsir93}, the final results are 
displayed in Table \ref{tab:remu} and compared with previous
calculations. For the pionic ratio, the previous predictions are
confirmed with better precision. The discrepancy in the predictions
for the kaonic ratio is mainly due to the fact that the asymptotic
behaviour of form factors in the model used in Ref.~\cite{finke96} is
incompatible with QCD. Comparing the theoretical with the present 
experimental values in (\ref{eq:remuexp}), there is room for
new physics to be detected with more accurate measurements.


\section{Nonleptonic $K$ decays}
\noindent 
CHPT has also had a big impact on nonleptonic $K$
decays. However, in contrast to semileptonic decays, already at NLO,
 $O(G_F p^4)$, not all LECs are known. Therefore, nonleptonic decays 
without any LECs at NLO have always been theorists' favourites.
They are unambiguously predicted to $O(G_F p^4)$ in terms of the two
known LECs of lowest order. Clearly, estimates of NNLO contributions
are needed for a meaningful comparison with experiment.  

Two early examples are the decays $K_S \to \gamma \gamma$
\cite{ksgg_chpt} and $K_L \to \pi^0 \gamma \gamma$ \cite{klpigg_chpt}
where recent experimental developments have greatly clarified the
situation. Estimates of higher-order corrections were made in two
directions. 
\begin{itemize} 
\item Rescattering corrections can be calculated in a largely model
  independent way from unitarity \cite{cdm93,cep93,kh94}. For $K_S \to
  \gamma \gamma$ these corrections can essentially be expressed in
  terms of the rates for $K \to \pi \pi$. On the other hand, the
  corrections are more involved for $K_L \to \pi^0 \gamma \gamma$ and
  they are sizable.
\item A comprehensive treatment of LECs
  of $O(G_F p^6)$ is beyond present technology. From experience with
  strong amplitudes, one  expects vector meson exchange to be
  important whenever vector mesons contribute at all. They
  cannot contribute to  $K_S \to \gamma \gamma$ but they could
  have a substantial influence on  $K_L \to \pi^0 \gamma
  \gamma$. Assuming that vector meson exchange is indeed dominating at 
  $O(G_F p^6)$, the contributing LECs of  $O(G_F p^6)$ can be
  parametrized by a single dimensionless constant $a_V$ \cite{cep93,dap97}. 
\end{itemize}    

\begin{figure}[ht]
    \vskip -.5 cm
    \begin{center}
    \begin{minipage}[t]{0.4\linewidth}
    \raisebox{48mm}{ 
\renewcommand{\arraystretch}{1.3}
\begin{tabular}{lcc}
      
        \hline \hline
       &   &   \\[-13pt] 
              &   & $B(K_S \to \gamma \gamma) \cdot 10^{6}$
                                                   \\[3pt]
          \hline  \\[-13pt]
         KLOE \cite{kloe_ksgg}       &               & $2.26(12)(06)$  \\[3pt] 
         CHPT \cite{ksgg_chpt} & & $2.15(20) $ \\[3pt] 
        \hline \hline
      \end{tabular} }
     \vskip -2.7 cm 
     \caption{Comparison of experimental results with the CHPT
     prediction for $K_S \to \gamma \gamma$ (courtesy of
     Matteo Martini).
     \label{fig:ksgg} }
    \end{minipage}
    \hspace*{\fill}
    \begin{minipage}[t]{0.55\linewidth}
\begin{flushright} 
    \includegraphics[width=0.9\linewidth]{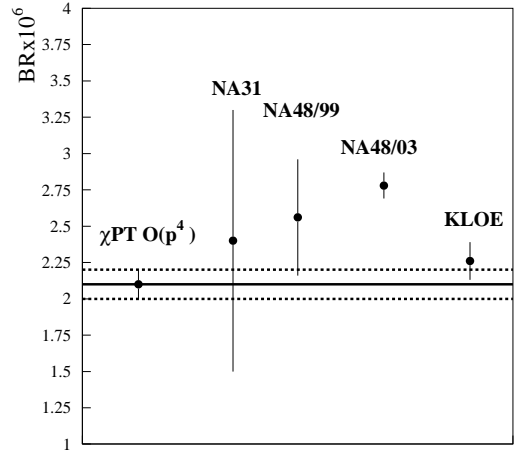}
\end{flushright} 
    \end{minipage}
   \end{center}
\end{figure}

\vskip -.4 cm 
Updating the original calculation \cite{ksgg_chpt} and assigning an
uncertainty of 10$\%$ to the rate for $K_S \to \gamma \gamma$ in view 
of the above discussion, the CHPT prediction for the branching ratio
is 
\begin{equation} 
B(K_S \to \gamma \gamma)|_{\rm CHPT} = (2.15 \pm 0.20) \cdot 10^{-6}~.
\label{eq:ksgg}
\end{equation}
As displayed in Fig.~\ref{fig:ksgg}, the first experimental results
were consistent with expectations within large errors but the last
precise measurement of NA48 \cite{na48_ksgg} was significantly bigger
than the CHPT branching ratio (\ref{eq:ksgg}). To my knowledge, no
serious attempt was made to find a mechanism that would raise the
chiral prediction (\ref{eq:ksgg}) substantially. The most recent precise
measurement from KLOE \cite{kloe_ksgg} shown in Fig.~\ref{fig:ksgg} is
therefore most welcome, reestablishing the agreement between theory
and experiment.

Turning to $K_L \to \pi^0 \gamma \gamma$, experimental data have
always been in agreement with the chiral prediction that the pion-loop
contribution dominates the two-photon spectrum. However, until very
recently there were conflicting experimental results for the decay
rate that is affected significantly by both rescattering corrections
and by higher-order LECs. With the final analysis of KTeV for  
$K_L \to \pi^0 \gamma \gamma$ released this year \cite{ktev_klpigg}, 
there is now excellent agreement between
theory and experiment for both rate and spectrum in terms of
the single parameter $a_V$. The final results
\begin{eqnarray} 
B(K_L \to \pi^0 \gamma \gamma) \cdot 10^6 &=& \left\{
\begin{array}{ll} 1.36 \pm 0.03 \pm 0.03 \pm 0.03 & \hspace*{.4cm} 
{\rm NA48} ~\cite{na48_klpigg} \\
1.29 \pm 0.03 \pm 0.05 & \hspace*{.4cm} {\rm KTeV} ~\cite{ktev_klpigg}
\end{array} \right. \\
a_V &=& \left\{
\begin{array}{ll} - 0.46 \pm 0.03 \pm 0.04 & \hspace*{1.2cm} 
{\rm NA48} ~\cite{na48_klpigg} \\
- 0.31 \pm 0.05 \pm 0.07 & \hspace*{1.2cm} 
{\rm KTeV} ~\cite{ktev_klpigg}
\end{array} \right.
\end{eqnarray} 
document that patience is sometimes appropriate for nonleptonic $K$
decays. An important consequence is that the CP conserving
contribution to $K_L \to \pi^0 e^+ e^-$ via $K_L \to \pi^0 \gamma^* 
\gamma^* \to \pi^0 e^+ e^-$ is definitely negligible compared
with the CP violating amplitudes.

It would be premature to conclude that higher-order corrections in
nonleptonic $K$ decays are under control in general but the situation
for $K_S \to \gamma \gamma$ and  $K_L \to \pi^0 \gamma \gamma$ is
certainly encouraging. More experimental results are already available or
forthcoming. A recent review of nonleptonic $K$ decays in CHPT can be
found in Ref.~\cite{prades07}.

\section{Other topics}
\noindent
Finally, I briefly review here a few other
interesting recent developments in CHPT.
\subsection{Radiative pion decay $\mbf{\pi \to e \nu_e \gamma}$}
\noindent 
Resonance contributions to the vector and axial-vector form factors
governing the structure dependent part of the $\pi \to e \nu_e \gamma$
amplitude were calculated in Ref.~\cite{mp07}. The relevant LECs of
$O(p^6)$ were also estimated in Ref.~\cite{up08} where in addition
radiative 
corrections for the process were performed in a CHPT framework. The
very recent PIBETA experiment \cite{pibeta08} finds no evidence for a
previously reported tensor contribution and is in agreement with
theoretical expectations.

\subsection{Chiral $\mbf{SU(2)}$ vs. $\mbf{SU(3)}$}
\noindent
In the limit where the strange quark mass $m_s$ is much bigger than
$m_u$, $m_d$ and all external momenta, chiral $SU(3)$ reduces to the
two-flavour case. Such a procedure allows to determine the 
$m_s$-dependence of LECs in chiral $SU(2)$ and it provides relations
between the LECs of the two- and three-flavour chiral Lagrangians. To
$O(p^4)$, such relations were established already in the classic paper
of Gasser and Leutwyler \cite{glsu3}. Recently, the relations have been
worked out to $O(p^6)$ \cite{ghis07}. The results are expected to be
useful for determining some of the LECs of $O(p^6)$ to
allow for an efficient comparison between theory and experiment to
two-loop accuracy. 

\subsection{$\mbf{\eta \to 3 \pi}$ decays}
\noindent
The decays $\eta \to 3 \pi$ are prominent examples for large chiral
corrections. A complete $O(p^6)$ calculation was performed recently by
Bijnens and Ghorbani \cite{bg07}. The corrections of $O(p^6)$ turn out
to be somewhat larger than previously obtained with dispersive
methods \cite{eta_disp}. However, a very recent measurement of both
$\eta \to 3 \pi^0$ and $\eta \to \pi^+ \pi^- \pi^0$ by KLOE
\cite{kloe_eta07} indicates that there are still discrepancies between
theory and experiment, 
especially in the slope parameters. Once again, it may be necessary to
have better theoretical control of the $O(p^6)$ LECs involved.

\section{Conclusions}
\noindent 
Thirty years after its conception, there is still significant progress
in CHPT along several lines. 

The impressive precision in pion pion scattering obtained by
combining CHPT with dispersion theory is now being challenged
experimentally, with data mainly from
$K_{e4}$ and $K \to 3 \pi$ decays.

Kaon physics is a traditional stronghold of CHPT. Even if some
issues remain to be clarified, $K_{l3}$ decays provide at present the
best source 
for extracting the CKM matrix element $V_{us}$. The recent calculation
of the ratios  $\Gamma (P \to e \nu_e)/\Gamma (P \to \mu \nu_\mu)$ to 
$O(e^2 p^4)$ with very small theoretical uncertainties constitute a
challenge for experimental confirmation or the possible detection of
new physics. The history of the nonleptonic decays  $K_S \to \gamma 
\gamma$ and $K_L \to \pi^0 \gamma \gamma$ suggests that
sometimes  patience is called for in this field.

In general, CHPT stands for precision physics at low energies in
several areas, allowing for significant tests of the Standard Model. In
particular, CHPT has established itself as the only reliable
method for isospin violating and electromagnetic corrections. Further
progress in the field will depend on progress in the determination of
LECs. 


\end{document}